# Hadron Physics with CLAS12

Stepan Stepanyan[a]

[a]*Jefferson Lab, 12000 Jefferson Avenue, Newport News, VA (USA)*

**Abstract.** Hadron spectroscopy has been an essential part of the physics program with the CLAS detector in experimental Hall B at Jefferson Lab. Production of baryon and meson resonances with high energy (polarized) electron and photon beams was studied on a veriety of targets, ranging from hydrogen to lead. Physics topics of interest include: investigation of the spectrum of baryon and meson resonances, transition form-factors, meson-nucleon couplings (mesons in nuclei), and search for exotic and missing states. With the 12 GeV upgrade of the CEBAF machine, hadron spectroscopy in Hall B will be extended to a new domain of higher mass resonances and the range of higher transferred momentum using up to 11 GeV electron beams and the upgraded CLAS12 detector. In this paper a brief description of the CLAS12 detector and the physics program adopted for 12 GeV with emphasis to baryon and meson spectroscopy is presented.

## CLAS12 DETECTOR

The 12-GeV upgrade of the CEBAF at Jefferson Lab includes upgrade of the machine energy to 12 GeV and of the equipment in the experimental Halls. The detector in Hall B, CLAS, will be upgraded to CLAS12 [1]. The CLAS12 detector will make major contributions in many areas of hadron physics. In particular, CLAS12 will have design features that are essential for probing the new physics of the Generalized Parton Distributions (GPDs). CLAS12 will carry out the core program for the study of the internal dynamics and 3D imaging of the nucleon, and quark hadronization processes. These studies are carried out by measuring exclusive and semi-inclusive processes, using hydrogen and nuclear polarized and unpolarized targets. CLAS12, see Figure 1, consists of two parts, forward detector (FD) and central detector (CD).

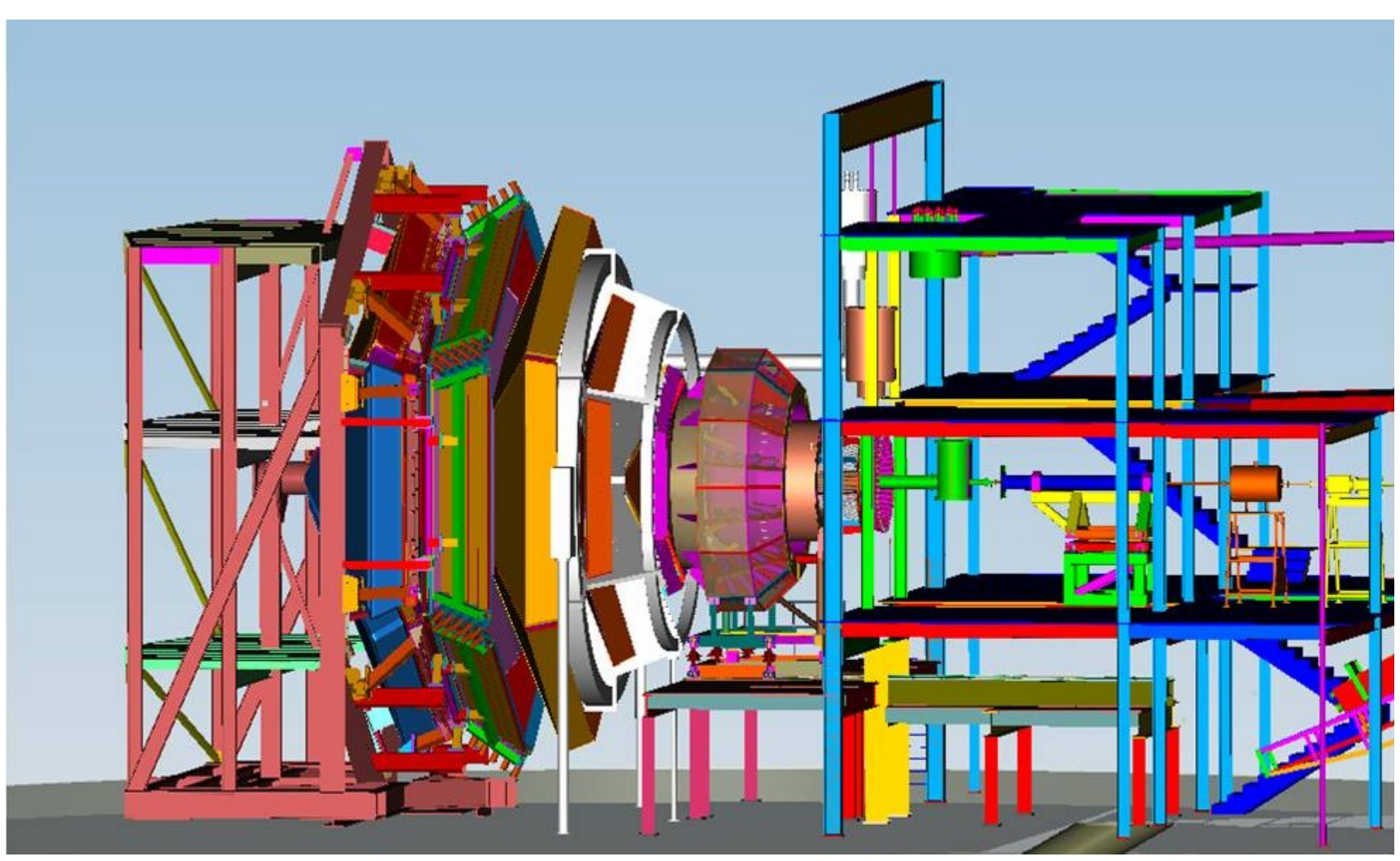

**FIGURE 1.** 3-D model of the CLAS12 detector in the experimental Hall B at JLAB.

FD makes use of many existing detector components of CLAS and the Hall B infrastructure. Major new components include the superconducting torus coils that now cover only the forward angle range, a new gas Cerenkov counter for electron/pion separation, a new forward vertex tracker and forward drift chambers, and additions to the electromagnetic calorimeters and time-of-flight system. The CD is based on a 5T superconducting solenoid magnet and consists of barrel tracker and a time-of-flight system. Design characteristics of CLAS12 are presented in Table 1. There are ongoing efforts in the collaboration to supplement the base equipment in Hall B with detector components that will improve overall CLAS12 capabilities (e.g. neutral detection in CD, high energy kaon identification in FD).

**TABLE (1).** Design characteristics of CLAS12.

| Parameters | Forward Detector | Central Detector |
|---|---|---|
| Charged tracks: | | |
| polar angular range ($\theta$) | $5^o$ to $35^o$ | $35^o$ to $125^o$ |
| polar angular resolution ($\delta\theta$) | < 1 mr | < 10 mr – 20 mr |
| azimuthal angular resolution ($\delta\varphi$) | < 4 mr | < 5 mr |
| momentum resolution ($\delta p/p$) | < 1% @ 5 GeV/c | < 5% @ 1.5 GeV/c |
| Neutral particles: | | NA |
| polar angular range ($\theta$) | $5^o$ to $40^o$ | |
| angular resolution | < 4 mr | |
| energy resolution ($\sigma E$) | $0.1/\sqrt{E}$ | |
| Particle identification: | | |
| $e/\pi$ | full range | --- |
| $\pi/p$ | full range | < 1.25 GeV/c |
| $K/\pi$ | full range | < 0.65 GeV/c |
| $K/p$ | < 4 GeV/c | < 1 GeV/c |

One of the key improvements for CLAS12 will be the increase in the operating luminosity to $10^{35}$ $cm^{-2}$ $sec^{-1}$. The limitation in luminosity is given by electromagnetic background generated in the first region of drift chambers (mostly due to Moller electrons). An effective shield for Moller electrons is a solenoid field around the target as has been shown during CLAS operations. The required luminosity will be achieved by improving the solenoid shield and by decreasing drift chamber cell sizes.

## Available Beams and Targets for CLAS12

Electron beam energies at the 12 GeV CEBAF machine will be in multiples of ~2.2 GeV. For Halls A, B, and C maximum available beam energy will be 11 GeV at 5 pass, Hall D will have maximum energy of 12 GeV, 5.5 pass. CLAS12 can run experiments with up to 11 GeV longitudinally polarized electron beams. The existing bremsstrahlung photon tagging system in Hall B [2] will not run with electron beams above 7 GeV. Currently there are no plans to upgrade the photon tagging system and therefore only tagged photons with 1, 2, or 3 pass electron beams will be available for CLAS12. A major part of the physics program for hadron spectroscopy with CLAS12 will require the study of photoproduction of baryon and meson resonances. There are two options for studying high energy photoproduction reaction with CLAS12: (a) experiments with electron scattering at very small angles (photon beams with small virtualities, $Q^2<0.05$ $[GeV]^2$), (b) experiments with untagged photon beams. In both cases, the electron beam will traverse the physics target. In the first case, scattered electrons will be detected in a specialized detector at polar angles from $2^o$ to $4^o$, while in the second case the scattered electron kinematics will be deduced from a missing momentum analysis after detecting the full hadronic final state in CLAS12.

CLAS12 will retain capability of running on variety of targets ranging from liquid hydrogen, deuterium, $^3$He and $^4$He to solid targets. High field uniformity ($\Delta B/B<10^{-4}$) in the target region of the 5T solenoid magnet will allow operation of dynamically polarized targets such as $NH_3$ and $HD_3$. The precision electron beams will allow use of thin gaseous targets, important for experiments where detection of low energy recoil particles is required.

## CLAS12 PHYSICS PRORAM

A broad physics program is adopted on CLAS12 using up to 11 GeV polarized electron beams and verity of targets. A more complete discussion of the physics program is given in [3]. This program includes

- Nucleon structure studies by mapping out nucleon GPDs and Transverse Momentum Distributions (TMDs) using exclusive and semi-inclusive processes with deeply inelastic electron scattering reactions
- Precision measurements of structure functions and forward parton distributions at high $x_B$
- Elastic and transition form factors at high momentum transfer
- Hadronization and color transparency
- Hadron spectroscopy, studying heavy baryons and mesons with ordinary and exotic quantum numbers

A major focus of CLAS12 will be to determine the Generalized Parton Distributions (GPDs) using deeply exclusive electroproduction reactions. Electron scattering has been used successfully to study the structure of the nucleon for decades. Elastic scattering and deeply inelastic scattering give us two orthogonal one-dimensional projections of the proton. Elastic scattering measures the probability of finding a proton with a transverse size $b_\perp$ matching the resolution of the probe given by the momentum transfer t: $b_\perp \sim 1/\sqrt{|t|}$. The expression relates the momentum transfer to the transverse size of the proton probed in the interaction. Deeply inelastic scattering probes the longitudinal momentum distribution of the quarks, but has no sensitivity to the transverse dimension. These two aspects are illustrated in the first two panels of Figure 2 [4]. The information resulting from these two types of experiments is disconnected, and does not allow us to construct the image of a real 3-dimensional proton.

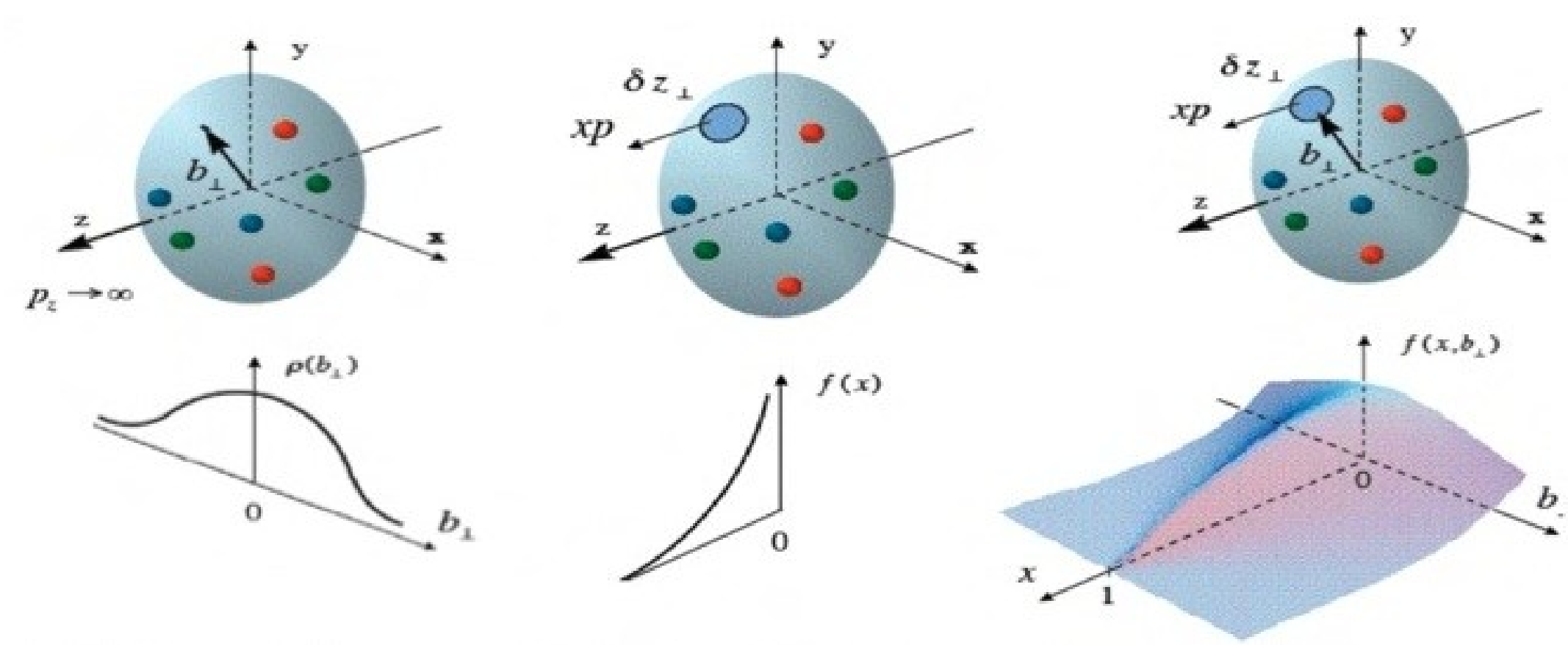

**FIGURE 2.** Representations of the proton properties probed in elastic scattering (left), deeply inelastic scattering (center), and deeply exclusive scattering processes (right).

Deeply exclusive scattering processes connect both transverse and longitudinal information including their correlations as described by GPDs. This is shown in the third panel of Figure 2. GPDs depend on three variables, $(x,\xi,t)$. The simplest process for accessing GPDs is deeply virtual Compton scattering (DVCS). Imaginary part of Compton amplitudes that are linear combinations of GPDs (at specific kinematic point, $x=\xi$) can be extracted from measurements of (beam/target) spin asymmetries. By measuring the total cross section of DVCS one will access the square of the real part of Compton amplitudes that contain integrals of GPDs over variable x. The real part of the Compton amplitude can be accessed directly in photoproduction of lepton pairs, so called time-like Compton scattering process. All these measurements can be done using the CLAS12 detector. Once the GPDs are measured they allow the construction of a 3-dimensional image (two in transverse space and one in longitudinal momentum) of the proton in what has been called ``nucleon-tomography" [5]. Moments of GPDs will allow the study of the mass and angular momentum distributions of quarks, and the forces and pressure distributions in the proton.

## HADRON SPECTROSCOPY WITH CLAS12

Spectroscopy of hadrons (mesons and baryons) is one of the key tools for studying the theory of strong interactions, Quantum Chromodynamics (QCD), in the non-perturbative regime (*i.e.*, confinement). CLAS already made significant contributions to hadron spectroscopy, see e.g. [6, 7, 8, 9, 10]. To date, a large amount of experimental data on electromagnetic production of mesons and baryons has been collected by CLAS. However, more data will be necessary to guide improvements in hadronic phenomenology and to compare with lattice QCD calculations.

The major part of the data obtained so far with CLAS is restricted to the lowest mass states formed with the lightest quarks: up, down and strange. A complete picture of QCD in the strong-coupling (non-perturbative) regime requires extension of hadron spectroscopy studies to higher masses and/or higher transferred momenta.

The key experiments in hadron spectroscopy that are planned for the upgraded CLAS12 detector will study reactions produced by both, quasi-real and virtual photons. They include:

- Photoproduction of high-mass mesonic states (consisting of ordinary mesons, hybrids, and mesons with exotic $J^{PC}$) using $LH_2$ and light nuclear targets;
- Higher mass baryon production, *e.g.*, $\Sigma^*$ and $\Xi^*$ baryons;
- Studies of N* resonances at the transition between long- and short-distance scales as resolved by virtual photons.

It is essential that the outlined program will *simultaneously* study both meson and baryon states, because the data analysis and physics aspects of these states are intrinsically related [11, 12]. Furthermore, it is also important that hadron production is studied in a wide range of scales (from high to almost zero $Q^2$).

The proposed technique for obtaining tagged (*quasi*-)real photons for hadron spectroscopy experiments is different from the one presently used in Hall B and different from the one planned for GlueX. We plan to use quasi-real photons produced by electrons scattered at very forward angles (*i.e.*, scattering angles <$4^o$). Small angle forward electron tagger, LowQ setup, will be used in coincidence with the detection of multi-particle final states with the CLAS12 detector to study electroproduction at $Q^2 < 0.05$ $GeV^2$. Electroproduction at these very small values of $Q^2$ using unpolarized electrons is equivalent to photoproduction using partially linearly polarized photons [13]. Availability of high intensity, high precision electron beams will allow us to achieve required luminosities on very thin targets (*i.e.*, gas targets) without increasing the rate of accidentals as would be the case for energy-tagget real photon bremsstrahlung beams. In turn, this will allow detection of low energy recoils (*e.g*., coherent scattering experiments) and spectators (*e.g.*, scattering off of the neutron in the deuteron). Knowledge of the photon linear polarization, high fluxes, together with the use of the nearly $4\pi$ coverage for hadronic final states of CLAS12, will allow the study of hadron spectroscopy in a competitive and complementary experimental environment to the already planned coherent bremsstrahlung production experiment in Hall-D.

## Meson Spectroscopy

A complete mapping of meson resonances in the mass region of 1 to 3 GeV will be particularly important for a better understanding of the QCD confinement mechanism. QCD predicts the existence of several new types of states beyond the constituent quark model: glueballs, hybrids, multi-quark states [14, 15]. Gluons play a central role in strongly interacting matter - quark confinement is due to gluonic forces. The clearest most fundamental experimental signature for the presence of dynamics of gluon degrees of freedom is the spectrum of gluonic excitations of hadrons. Gluonic excitations of mesons with ``exotic" quantum numbers, *i.e.*, quantum numbers not accessible to the qq-bar system, would be the most direct evidence for these states. Determining the properties of such states would shed light on the underlying dynamics of quark confinement.

The identification of these states has been difficult, as high mass resonances are generally broad and overlapping, and often have similar quantum numbers (mixing). To determine meson quantum numbers, partial wave analysis (PWA) is conducted (in a broad sense, fits to the angular distributions of final states). A complete PWA requires high event statistics, as well as high resolution and large geometrical acceptance of the detector. Meson spectroscopy at CLAS12, using the LowQ tagger, and coherent production on light nuclei will fulfill many of these stipulations.

### *Meson Spectroscopy in Coherent Production on Light Nuclei*

In the electromagnetic production of t-channel meson resonances at moderate energies the main physics background arises from associated production of baryon resonances that decay into the same final state particles. Often these particles in both production reactions occupy the same phase space (see Figure 3), and therefore, it becomes impossible to separate both processes using kinematic cuts. The contribution of baryon resonances to the final state complicates PWA analysis. Coherent production of meson resonances on nuclear targets, when the recoil nucleus remains intact, is a clean way to eliminate baryon resonances. A particular case of such processes is the coherent production off of light nuclei, *e.g.*, $^3$H, $^3$He, $^4$He. In these case the recoil nucleus can be detected in order to ensure that it remains intact.

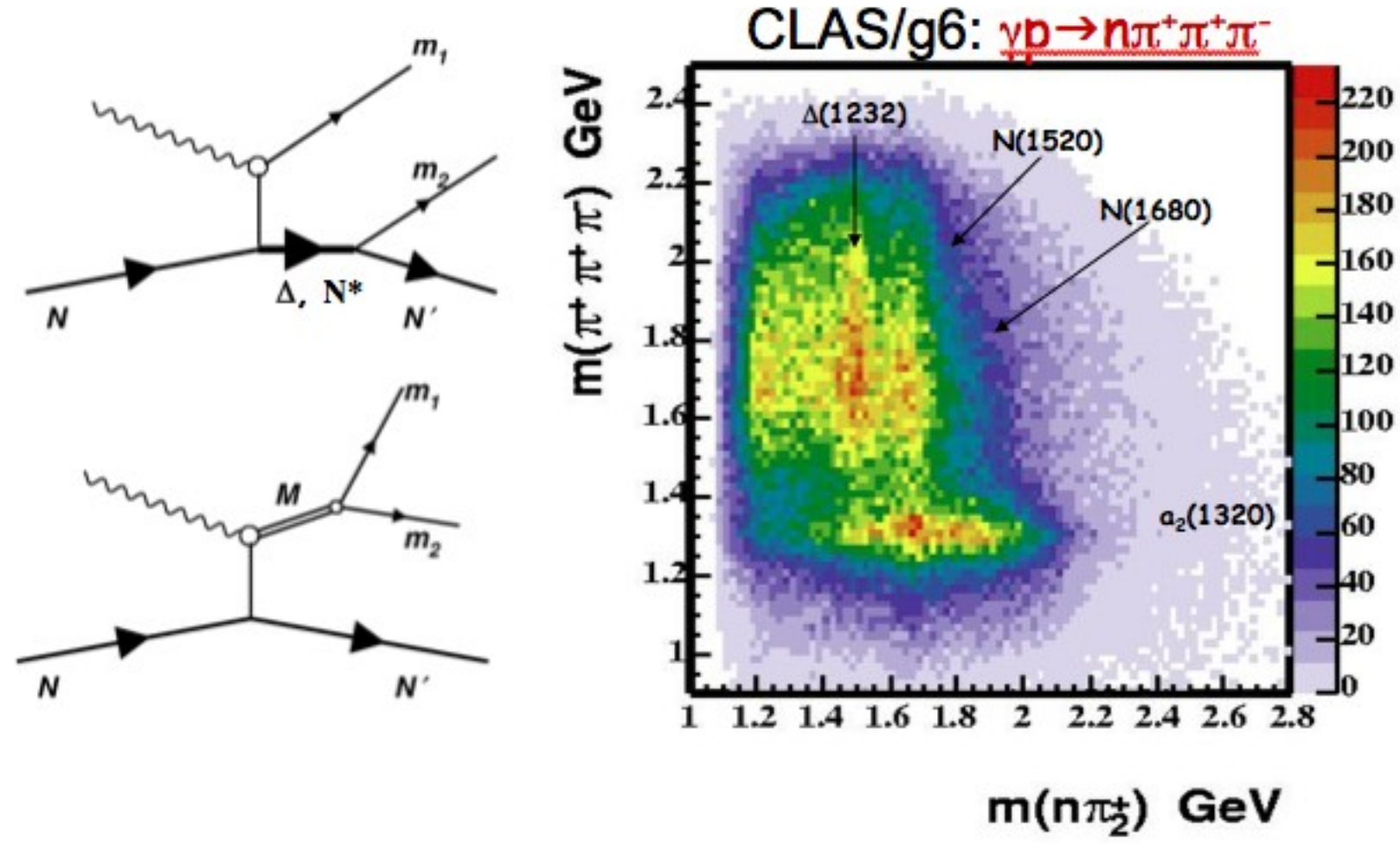

**FIGURE 3.** Mixing of meson and baryon resonances with the same final state particles. 2-D distribution is from analysis of 3-π final states using CLAS photoproduction data.

For meson masses from 1 to 3 GeV/c$^2$, the minimum transferred momentum, $t_{min}$, at beam energies up to 10 GeV, ranges from 0.02 to 0.2 (GeV/c)$^2$. At these transferred momenta, reduction of yields due to the nucleus form factor is expected to be only a factor of a few, while the energy obtained by the recoiling nucleus, 5 to 30 MeV for $^3$H, $^3$He, and $^4$He, will be enough to study them using thin gas targets and lo density tracking detectors. Such experiments can only be conducted with high intensity electron beams, where the low target density is compensated by the high flux of quasi-real photons. High precision electron beam, with sub-millimeter cross section, will allow to use small diameter target cell with thin target cell walls at high pressure, a critical component for the detection of low energy recoiling particles.

Besides the elimination of baryon resonances, coherent production on nuclei has other advantages as well. In many cases it imposes constrains on the allowed helicity states of the produced meson, and on the possible exchange particles. These will significantly aid the PWA.

Examples of such reactions are the coherent production of $\pi\eta$ and $\pi\eta'$ final states on $^4$He [16]. The attractive feature of these final states is that in P-wave they have exotic quantum numbers, $J^{PC}=1^{-+}$. Photoproduction of $\pi\eta$ and $\pi\eta'$ on the nucleon proceeds only via C-odd $\rho$ or $\omega$ exchanges. Since $^4$He has isospin-0, only $\omega$ the natural parity exchange is allowed. Due to spin-0 of $^4$He, the helicity of the final state (at small angles) will be equal to the helicity of the incoming photon (SCHC). This will significantly simplify PWA.

Recently CLAS completed such an experiment. The experiment used the 6 GeV electron beam incident on the pressurized $^4$He gas target. The recoil helium nucleus was detected in a Radial Time Projection Chamber (RTPC) based on cylindrical GEMs. With 6 GeV beam, the mass range < 2 GeV of coherently produced meson will be studied. With CLAS12 and a 11 GeV electron beams this mass range can be extended up to 3 GeV.

### *Meson Spectroscopy with Electroproduction at Very Small $Q^2$*

The general idea of PWA is to parameterize the intensity distribution in the space of quantum numbers available to the observed final states. The intensity distribution is written as a sum of interfering and non-interfering amplitudes (partial waves). A maximum likelihood fit is done to the intensity distribution by a set of given partial waves and reasonable assumptions of the production mechanisms. The goodness of the fit is related to the statistics and the rank of the production matrix, and the number of parameters to be fitted.

The knowledge of photon polarization simplifies the PWA by giving direct information on the production mechanisms and therefore, reducing the rank of the fit. In electroproduction at very low $Q^2$, we will be able to measure, in an event-by-event basis, the linear polarization of the photons.

The setup needed for electron detection at very small angles is not in the base design of the CLAS12. However, collaboration is already started efforts for the design and construction of a device that will allow detection of scattered electrons in the angular range from $2^o$ to $4^o$. A concept of the LowQ detector is presented in Figure 4, showing a setup with tracking device and electromagnetic calorimeter. Tracking is important for precise measurements of scattered electron angles, which is important for determination of the virtual photon polarization vector. The calorimeter will be used to measure the scattered electron energy and for triggering.

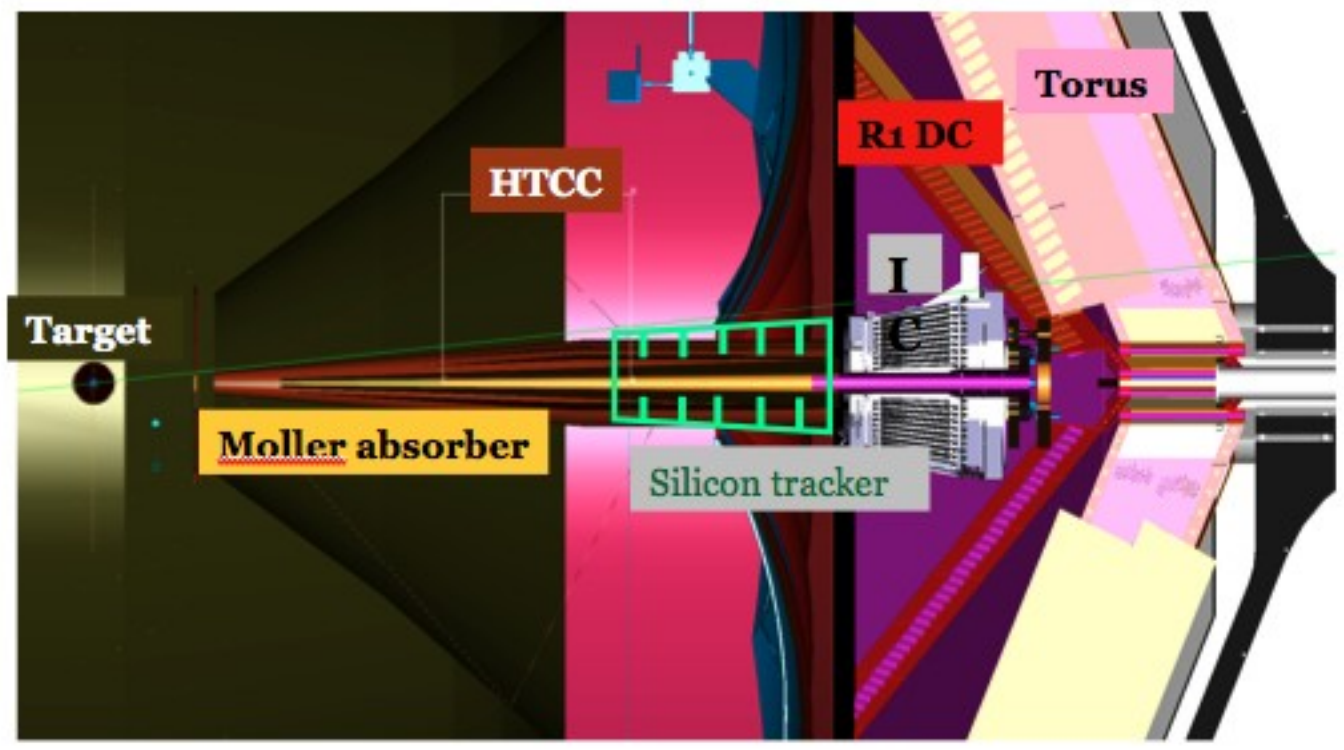

**FIGURE 4.** CLAS12 inner bore with proposed LowQ setup.

Spectroscopy studies of mesons have started at JLab with CLAS at lower energies. Preliminary results of these experiments show the viability of such studies using the current CLAS configuration. PWA of simple final states ($\pi\pi\pi$) have already been carried out successfully using current CLAS data [7]. However, with the limited acceptance of the present CLAS and only circular polarization of the photon beam, there are many analysis ambiguities created by the assumptions on the production mechanisms and the baryon backgrounds. These ambiguities will be mostly resolved when using linearly polarized high energy (greater than 8 GeV) photons and larger acceptances. At higher energies we will be able to map out the mesons in the interesting 1 to 3 GeV mass range and, most importantly, to better kinematically differentiate mesons from baryons. More specifically, we plan to study mesons decaying into three or four final states (for example: $\rho\pi$, $\eta\pi$, $\phi\eta$, $b_1\pi$, KK*, ...). A Letter-of-Intend for such measurements with a proposal to design and build a new LowQ setup was submitted to JLAB PAC35 [17].

## Baryon spectroscopy

Ground and excited baryon states provide a wealth of information on non-perturbative QCD in addition to that available using mesons. First, baryons are the simplest systems that manifest the non-abelian nature of QCD. This results in a complex internal structure whose degrees of freedom and dynamics may depend on the distance scale probed. Second, the predicted mass spectrum involves many transitions to a variety of spin-flavor multiplets (56, 70 and 20-plets), which partially overlap in excitation energy. Many of these states have not yet been seen. Understanding and unraveling this rich structure requires experimental data of high precision using a variety of exclusive channels, final hadronic state invariant masses and photon virtualities. The enhanced kinematic range available at 12 GeV will make feasible a study of the evolution of baryon structure and quark binding mechanisms in the transition between QCD strong coupling and asymptotic freedom.

### *Cascade Baryons*

The double-strangeness cascades have several advantages when it comes to baryon spectroscopy. First, two of the three valence quarks are heavier ($m_s$ ~ 100 MeV) than light quarks, which reduces the uncertainties in extrapolations of lattice gauge calculations for the cascade mass. Second, the width of the excided cascades are typically a factor of ten smaller than their N* counterparts. Third, the detached decay vertex for many cascades allows experiments to more easily separate cascades from various backgrounds. These advantages can only be utilized if there is sufficient beam energy to produce the Ξ's in reasonable quantities.

The Particle Data Group [18] shows that there are many excited Ξ*'s with fairly narrow widths based on older bubble-chamber or hadronic-beam experiments. In some cases, the data for Ξ*'s are not very consistent, e.g. some experiments that have reported a Ξ* should have seen other Ξ*'s in their mass window, but did not. Moreover, of the 22 Ξ candidates expected in the SU(3) multiplets, only six are well-established. Besides the spectroscopy of cascade states, the narrow widths of ground and excited states will enable study of isospin violating mass splitting, that is not feasible in the N* sector where states are around 100 MeV wide.

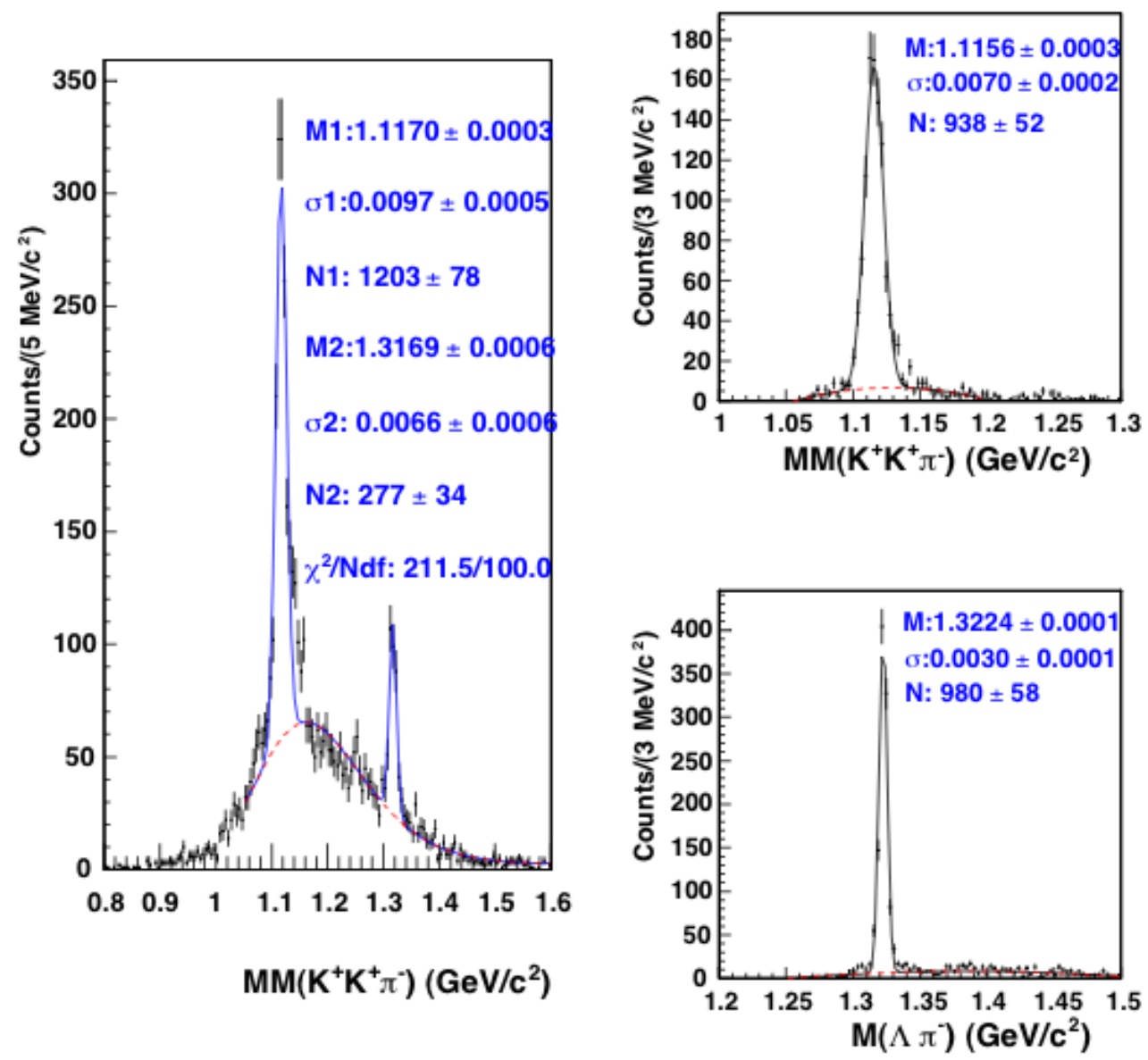

**FIGURE 5.** Left: ($K^+K^+\pi^-$) missing mass spectrum. The dashed background shape is obtained from events with an additional $\pi^+$ in the same event; Top right: ($K^+K^+\pi^-$) missing mass with a 3σ cut on the $\Xi^-$ region (in the ($K^+K^+$) missing mass); Bottom right: ($\Xi^-$) invariant mass with a 3σ cut on the Λ region (in the ($K^+K^+\pi^-$) missing mass).

Experiments at the CLAS detector have shown that the ground state Ξ and the first excited Ξ* can be clearly seen in photoproduction from the proton. In ref. [9] the cascade resonances has been investigated in the reactions $\gamma p \rightarrow K^+K^+X$ and $\gamma p \rightarrow K^+K^+\pi^-X$. The mass splitting of the ground state ($\Xi^-$, $\Xi^0$) doublet is measured to be 5.4 ± 1.8 MeV/$c^2$, see Figure 5. Differential cross sections of the $\Xi^-$ are consistent with a production mechanism through intermediate hyperon state. With the 11 GeV electron beam and with excellent vertex and momentum/angular resolution, CLAS12 is the ideal place for studying spectrum and production dynamics of double-strangeness cascade baryons.

### *Excited nucleon states at 12 GeV*

The goal of the program for study of nucleon resonances at high energies is determine electro-couplings of prominent excited nucleon states (N*, Δ*) in the range $Q^2$ = 5 - 12 GeV$^2$. Experimental tools are differential cross sections and polarization observables in single and double pseudo-scalar meson production: $\pi^+n$, $\pi^0p$, ηp and $\pi^+\pi^-p$ measured in full polar and azimuthal angle ranges [19]. These data will, for the first time, allow us to study the structure of excited nucleon states in domains where the dressed quarks are major active degrees of freedom, and explore their emergence from QCD.

## SUMMARY

A broad physics program adopted at CLAS12 covers topics of nucleon and nuclear structure, formfactors and spin structure functions, hadron spectroscopy and hadronization. With available high precision electron beams, the CLAS12 detector in Hall B is well suited to carry out a rigorous program in hadron spectroscopy using (quasi) real and virtual photoproduction reactions. In particular, meson spectroscopy using electroproduction at small angles and coherent production on nuclei can be done only with CLAS12. Excellent momentum, angular, and vertex resolutions combined with large acceptance and excellent PID, makes CLAS12 a unique place to study long-lived hyperons, such as $\Xi^*$. The hadron spectroscopy program for CLAS12 is still under development. Shaping up the physics program in terms of physics proposals to JLAB PAC is just started. One proposal is already approved, and several LOI will be presented to PAC 35.

## ACKNOWLEDGMENTS

Authored by Jefferson Science Associates, LLC under U.S. DOE Contract No. DE-AC05-06OR23177. The U.S. Government retains a non-exclusive, paid-up, irrevocable, world-wide license to publish or reproduce this manuscript for U.S. Government purposes.

## REFERENCES

1. *CLAS12 Technical Design Report V5.1*, JLAB (2008).
2. D. Sober et al., *Nuclear Instrum. Methods A* **440**, 263 (2000).
3. V. Burkert, *arxiv:* **0810.4718** (2008).
4. A. Belitsky and D. Muller, *Nucl. Phys A* **711**, 118 (2002).
5. M. Burkhart, *Nucl.Phys. A* **711**, pp. 127-132 (2002).
6. M. Battaglieri et al. (The CLAS Collaboration). *Phys. Rev. Lett.* **102**, 102001 (2009)
7. M. Nozar et al. (The CLAS Collaboration). *Phys. Rev. Lett.* **102**, 102002 (2009)
8. I. G. Aznauryan et al. (The CLAS Collaboration), *Phys. Rev. C* **78**, 045209 (2008), *Phys. Rev. C* **80**, 055203 (2009).
9. L. Guo et al. (The CLAS Collaboration), *Phys. Rev. C* **76**, 025208 (2007).
10. B. McKinnon et al. (The CLAS Collaboration), *Phys. Rev. Lett.* **96**, 212001
11. F.E. Close, *AIP Conf. Proc.* **717**, 919 (2004).
12. T. Barnes, N. Black and P.R. Page, *Phys. Rev. D* **68**, 054014D (2003).
13. N. Dombey, *Rev. Mod. Phys.* **41**, 236 (1969).
14. N. Isgur and J. Paton, *Phys. Rev. D* **31**, 2910 (1985).
15. N. Isgur, R. Kokoski, and J. Paton, *Phys. Rev. Lett.* **54**, 869 (1985).
16. S. Stepanyan et al., *JLAB Experiment* **E-07-009** (2007).
17. M. Battaglier et al., *Letter-of-Intend to JLAB PAC 35*, **LOI-10-001** (2010)
18. Review of Particle Physics, *J. Phys. G: Nucl. Part. Phys.* **33** (2006).
19. R. Gothe et al., *JLAB Proposal to PAC 34*, **PR-09-003** (2009).